\documentclass[aps,10pt,prb,twocolumn,superscriptaddress]{revtex4-2}
\usepackage{graphicx}
\usepackage{latexsym}
\usepackage{amssymb}
\usepackage{amsmath}
\usepackage{amsfonts}
\usepackage[vcentermath]{youngtab}
\usepackage{upgreek}
\usepackage{bm,dsfont}
\usepackage{multirow}
\usepackage{enumitem}
\usepackage{color}
\usepackage{hyperref}
\hypersetup{
colorlinks = true,
linkcolor = [rgb]{0.70,0.13,0.13},
citecolor = [rgb]{0.13,0.55,0.13},
urlcolor = [rgb]{0.25, 0.41, 0.88}}
\usepackage{makecell}
\usepackage{comment}
\usepackage{verbatim}

\newcommand{\expval}[1]{\langle #1 \rangle}

\newcommand{\e}{\mathrm{e}}

\newcommand{\id}{\mathds{1}}
\newcommand{\E}{\mathop{\mathbb{E}}}

\newcommand{\dsZ}{\mathbb{Z}}

\newcommand{\scV}{\mathcal{V}}

\newcommand{\Tr}{\operatorname{Tr}}

\newcommand{\vect}[1]{{\bm{#1}}}

\newcommand{\eqnref}[1]{Eq.\,\eqref{#1}}
\newcommand{\figref}[1]{Fig.\,\ref{#1}}

\newcommand{\refcite}[1]{Ref.\,\onlinecite{#1}}

\begin{document}

\title{ShadowGPT: Learning to Solve Quantum Many-Body Problems\\ from Randomized Measurements}

\author{Jian Yao}
\affiliation{Department of Physics, Southern University of Science and Technology, Shenzhen, Guangdong 518055, China}
\affiliation{Department of Physics, University of California at San Diego, La Jolla, CA 92093, USA}
\author{Yi-Zhuang You}
\email{yzyou@physics.ucsd.edu}
\affiliation{Department of Physics, University of California at San Diego, La Jolla, CA 92093, USA}
\date{\today}

\begin{abstract}
We propose ShadowGPT, a novel approach for solving quantum many-body problems by learning from randomized measurement data collected from quantum experiments. The model is a generative pretrained transformer (GPT) trained on simulated classical shadow data of ground states of quantum Hamiltonians, obtained through randomized Pauli measurements. Once trained, the model can predict a range of ground state properties across the Hamiltonian parameter space. We demonstrate its effectiveness on the transverse-field Ising model and the $\mathbb{Z}_2 \times \mathbb{Z}_2$ cluster-Ising model, accurately predicting ground state energy, correlation functions, and entanglement entropy. This approach highlights the potential of combining quantum data with classical machine learning to address complex quantum many-body challenges.
\end{abstract}
\maketitle

\textit{Introduction.} --- The quantum many-body problem, such as predicting the ground state properties of a given many-body Hamiltonian, is a central challenge in physics. Traditional numerical methods\cite{methods_dmrg,methods_tensornetwork,Verstraete2004Rcond-mat/0407066,method_mc_RevModPhys.73.33,Blankenbecler1981M,Sandvik1999S,mlmthods_nn,mlmethods_Carrasquilla2017,Pfau2020A} have been developed to tackle these problems from first principles, leveraging fundamental quantum mechanics principles to calculate results from scratch. These methods have achieved remarkable success within their specific application domains. However, challenges remains in general cases due to the inherent complexity of representing \textit{generic} quantum many-body states on classical computers. 

Recent advancements in quantum computation, especially with the development of quantum devices and simulators, have illuminated an exciting avenue for tackling quantum many-body problems: using quantum computers to (approximately) prepare ground states of given Hamiltonians, using techniques such as adiabatic state preparation\cite{Whitfield2011S1001.3855,Albash2018A} or variational quantum eigensolvers\cite{Peruzzo2014A1304.3061,McClean2016T1509.04279,Kandala2017H1704.05018,Tilly2022T2111.05176}, and then study their physical properties by measurements. Nevertheless, operating these quantum algorithms demands expensive resources, making quantum experimental data highly valuable yet underutilized in terms of its full potential.

Recent study\cite{Huang2021P2106.12627} has shown a provable advantage in incorporating quantum data into classical computation. Some quantum many-body problems, challenging or even intractable for classical algorithms alone, can become manageable when augmented by quantum experimental data. This points to a fundamental gap between \textit{simulability} and \textit{learnability}—certain quantum systems, while difficult to simulate classically, can have their behaviors efficiently predicted by classical machine learning trained on quantum data. 

This work aims to explore the possibility of learning from randomized measurement data collected from quantum experiments to build classical predictive model for solving quantum many-body problems. Of course, it should be acknowledged that classical machine learning has its limitations and cannot resolve every aspect of quantum systems. For example, quantum chaotic systems are conjectured to be \textit{unlearnable}\cite{Garratt2023M2207.09476,Garratt2024P2305.20092,tf_Zhang_2024}, under the intuition that chaotic behavior is unpredictable, and the learnability transition\cite{Yuezhen-Niu2020L2010.11983,Ippoliti2024L2307.15011} has became an interesting topic of research.

\begin{figure}[t]
\centering
\includegraphics[width=0.45\textwidth]{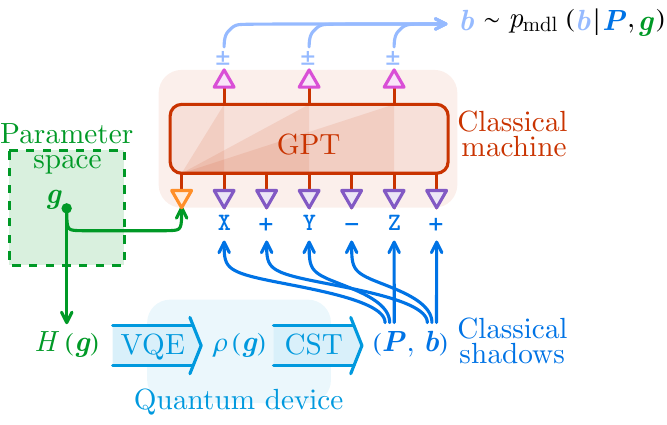}
\caption{Overview of the ShadowGPT approach. Given a Hamiltonian $H(\vect{g})$ parametrized by $\vect{g}$, a quantum device prepares the ground state $\rho(\vect{g})$ by variational quantum eigensolver (VQE), and measures it by classical shadow tomography (CST) to obtain the shadow data $(\vect{P}, \vect{b})$, where $\vect{P}$ is the sequence of random Pauli observables and $\vect{b}$ is the sequence of measurement outcomes. This data, $(\vect{g}, \vect{P}, \vect{b})$, is then used to train a GPT model to estimate the conditional distribution $p_\text{mdl}(\vect{b}|\vect{P}, \vect{g})$, enabling classical machine to generate classical shadows that mimics the quantum device, effectively solving quantum many-body problems in the parameter space.}
\label{fig:flow}
\end{figure}

Nevertheless, in this work, we will focus on classically learnable problems\cite{Lewis2024I2301.13169}. We assume access to quantum experiments capable of preparing ground states for a family of Hamiltonians and performing \textit{randomized measurements}\cite{Flammia2012Q1205.2300,Guta2018F1809.11162,Elben2023_random_toolbox} on such states. With this setup, can we construct a classical learning algorithm that, based on these quantum-generated data, can predict ground state properties for this family of Hamiltonians?

Previous studies, such as those in \refcite{Huang2021P2106.12627}, have addressed similar questions but focused on a discriminative task—predicting quantum phase labels from \textit{classical shadow}\cite{Huang2020_predict_properties} representations of quantum states. Our approach diverges from this by tackling a more challenging generative task, akin to an inverse problem: generating the classical shadow data of an underlying ground state when given the parameters of the parent Hamiltonian.

This generative task closely parallels tasks in natural language processing (NLP), specifically in generative modeling of sequential data, making Generative Pre-trained Transformer (GPT)\cite{tf_PhysRevA.104.032610,tf_Cha_2022,OpenAI2023G2303.08774,Zhang2023A2303.11717,Gozalo-Brizuela2023C2301.04655,Cao2023q2307.09025,tf_PhysRevB.107.075147,tf_PhysRevLett.130.236401,Fitzek2024R2405.21052,Nakaji2024T2401.09253,tf_lange2024transformer,tf_Sprague2024,tf_viteritti2024transformer,tf_wang2024predicting,tf_Zhang_2024}, a natural candidate for this context. We developed a model, ShadowGPT, trained on simulated ground state classical shadow data for a range of quantum Hamiltonians, as illustrated in \figref{fig:flow}. Once trained, ShadowGPT is capable of extrapolating to unseen Hamiltonians outside its training set, predicting ground state properties via classical shadow tomography. Our model demonstrates solid performance across various predictive tasks, including ground state energy, correlation functions, and entanglement entropy.

As a proof-of-principle theoretical study, we generated our training data through simulation rather than quantum experimentation, so our examples do not fully demonstrate the unique advantages of learning from quantum data. However, our approach is extendable in principle to scenarios where simulations are impractical or inefficient, thereby allowing learning to realize its full potential in future quantum-enhanced applications.

\textit{Task Description.} --- We consider a family of Hamiltonians $H(\vect{g})$ of an $N$-qubit quantum system, parameterized by a set of parameters $\vect{g}=(g_1,g_2,\cdots)$, whose specific form will be examplified later. Let $\rho(\vect{g})=\lim_{\beta\to\infty}\e^{-\beta H(\vect{g})}/\Tr\e^{-\beta H(\vect{g})}$ be the ground state of the Hamiltonian $H(\vect{g})$. We assume access to a quantum device capable of repeatedly preparing $\rho(\vect{g})$ for any specified parameter $\vect{g}$ and performing measurements on designated observables. Our goal is to use a limited number of quantum experiments (selecting only a few parameter points, with limited samples of quantum states and amounts of measurements) to collect data and train a model on a classical computer that predicts various physical properties of the ground state $\rho(\vect{g})$ across the parameter space.

Our challenge lies in predicting general properties of $\rho(\vect{g})$ without prior knowledge of which properties will ultimately be queried. In the worst-case scenario, this would require learning a full mapping from $\vect{g}$ to $\rho(\vect{g})$, which becomes impractical as the complexity of the density matrix scales exponentially with system size. A major breakthrough in recent years, known as classical shadow tomography\cite{Huang2020_predict_properties,Huang2022L}, addresses this issue by utilizing randomized measurement schemes to efficiently learn multiple low-complexity properties of quantum states in a scalable manner. These advances lay the foundation for our proposed approach.

\textit{Classical Shadow Representation.} --- Consider the \textit{randomized Pauli} measurement protocol\cite{Huang2020_predict_properties}, the simplest form of classical shadow tomography, starting with an $N$-qubit quantum state $\rho$ prepared on a quantum device. Each qubit is independently measured in a randomly chosen {Pauli observable} $P_i$ (for $i=1,2,\cdots,N$), where each observable $P_i$ is selected from ${X, Y, Z}$ with equal probability. Let the sequence of chosen Pauli observables be denoted as $\vect{P}=(P_1,P_2,\cdots,P_N)$. This measurement produces a corresponding sequence of outcomes $\vect{b}=(b_1,b_2,\cdots,b_N)$, where each binary variable $b_i = \pm 1$ represents the {outcome} of measuring observable $P_i$ on the $i$th qubit. The resulting pair $(\vect{P}, \vect{b})$ forms a measurement record, or a \textit{classical shadow} of the state $\rho$.

By performing repeated randomized Pauli measurements, we sample a distribution of classical shadows $(\vect{P}, \vect{b})$ with probability $p(\vect{b}|\vect{P},\rho)p(\vect{P})$ from the state $\rho$, where $p(\vect{P})=3^{-N}$ represents the uniform distribution of Pauli observables, and $p(\vect{b}|\vect{P},\rho)=\Tr\big(\rho \bigotimes_i\frac{1}{2}(\id+b_iP_i)\big)$ is simply the Born's rule of quantum mechanics. Given the ensemble of classical shadows, the original state $\rho$ can be reconstructed using the formula\cite{Huang2020_predict_properties}:
\begin{equation}\label{eq: shadow_tomo}
\rho=\E_{(\vect{P},\vect{b})}\bigotimes_i\frac{\id+3 b_i P_i}{2},
\end{equation}
where $\E_{(\vect{P},\vect{b})}$ denotes the ensemble average of classical shadows. This reconstruction formula enables estimation of various physical properties of $\rho$ efficiently, including both linear and non-linear observables, without the exponential overhead of reconstructing $\rho$ explicitly. For example, $\langle Z_iZ_j\rangle=\Tr\rho Z_iZ_j=\E_{(\vect{P},\vect{b})}b_ib_j\delta_{P_i=Z}\delta_{P_j=Z}$ can be estimated directly from the data correlation. For further information, we refer to reviews \refcite{Elben2023_random_toolbox,Huang2022L}.

Classical shadow tomography offers distinct advantages by providing classical data, in the form of $(\vect{P},\vect{b})$ pairs, that can be directly collected from quantum experiments; unlike the density matrix $\rho$, which can only be inferred indirectly. By representing quantum states through classical shadows, this method circumvents the exponentially complex task of reconstructing $\rho$, allowing a scalable and efficient approach to represent quantum many-body states on classical computers, which sets the stage for solving quantum many-body problems.

\textit{Data Collection and Tokenization.} --- In our data collection process, we use a classical computer to simulate a quantum device in preparing ground states $\rho(\vect{g})$ of the Hamiltonian $H(\vect{g})$ at selected parameter values $\vect{g}$ and gathering their classical shadows $(\vect{P}, \vect{b}) \sim p(\vect{b}|\vect{P}, \rho(\vect{g})) p(\vect{P})$ from randomized Pauli measurements. We envision that the data collection process can be directly implemented on realistic quantum hardware in the future. The data are stored in the format $(\vect{g}, \vect{P}, \vect{b})$, with the data distribution denoted as $p_\text{dat}(\vect{g}, \vect{P}, \vect{b})$.

For each triple $(\vect{g},\vect{P},\vect{b})$, the parameters $\vect{g}$ are a set of real numbers, and the observables $\vect{P}$ and measurement outcomes $\vect{b}$ are sequences that can be tokenized. It is sufficient to introduce the following vocabulary set $\scV=\{+1, -1, X, Y, Z\}$ to encode both the measurement outcomes $b_i=\pm1$ and Pauli observables $P_i=X,Y,Z$.

\textit{Modeling Strategy.} --- To model this data, we construct a sequence generative model (whose design will be specified later) to represent the conditional distribution $p_\text{mdl}(\vect{b}|\vect{P}, \vect{g})$, which probabilistically predicts measurement outcomes $\vect{b}$ of Pauli observables $\vect{P}$ on the ground state at the given Hamiltonian parameters $\vect{g}$. The model can be trained by minimizing the negative log-likelihood over the training data:
\begin{equation}\label{eq: loss}
\mathcal{L} = -\E_{(\vect{g}, \vect{P}, \vect{b}) \sim p_\text{dat}} \log p_\text{mdl}(\vect{b}|\vect{P}, \vect{g}).
\end{equation}

After training, the model can generate classical shadows on demand: given a parameter $\vect{g}$ and a uniformly sampled sequence of observables $\vect{P}$, it produces the corresponding measurement outcomes $\vect{b}$. From an ensemble of such $(\vect{P}, \vect{b})$ data, one can then infer the ground state properties from \eqnref{eq: shadow_tomo} for the specified parameter $\vect{g}$. This approach bypasses the exponential complexity to explicitly compute $\rho(\vect{g})$ from $H(\vect{g})$ by instead using a generative model that directly learns from quantum experiments to reproduce the classical shadow data of the ground state in response to the Hamiltonian parameter $\vect{g}$, effectively solving the many-body problem.

\textit{Example Hamiltonians.} --- To demonstrate the above methodology, we consider the following two test Hamiltonians. The first one is the one-dimensional transverse-field Ising model:
\begin{equation}
H = -\sum_{i=1}^{N}(1-g) Z_i Z_{i+1} - g \sum_{i=1}^N X_i,
\end{equation}
where $X_i$ and $Z_i$ are Pauli operators, with $Z_{N+i} = Z_i$ under periodic boundary condition. The transverse-field Ising model is driven by one parameter $g\in[0,1]$, and exhibits a quantum phase transition at $g=1/2$, such that the global $\dsZ_2$ spin flip symmetry $\prod_i X_i$ is spontaneously broken in the ferromagnetic phase ($g<1/2$) and restored in the paramagnetic phase ($g>1/2$). The second model is the one-dimensional $\dsZ_2\times\dsZ_2$ cluster-Ising model:
\begin{equation}
H = -g_1 \sum_{i=1}^{N}Z_{i-1} Z_{i+1} - g_2 \sum_{i=1}^{N} X_i - g_3 \sum_{i=1}^{N} Z_{i-1} X_{i} Z_{i+1},
\end{equation}
parametrized by $\vect{g}=(g_1,g_2,g_3)$ with $g_1,g_2,g_3>0$ and $g_1+g_2+g_3=1$. Periodic boundary condition is also assumed. The model has a global $\dsZ_2\times\dsZ_2$ symmetry corresponding to spin flips on odd and even sites, i.e., $\prod_{i\in\text{odd}}X_i$ and $\prod_{i\in\text{even}}X_i$.  It exhibits three phases: a spontaneous symmetry breaking (SSB) phase driven by $g_1$, a symmetric trivial phase driven by $g_2$ , and a symmetry-protected topological (SPT)\cite{Gu2009T0903.1069,Pollmann2012S0909.4059,Son2012T1111.7173} phase driven by $g_3$. The three phases are cyclically connected by a triality transformation\cite{Lu2024E2406.12151,Lu2024R2405.14939}, making the phase diagram invariant under the permutation of $g_1,g_2,g_3$.

The motivation to study these two models is twofold. First, they are well-understood models with known phase diagrams and physical properties, providing a benchmark for our model's performance. Second, ground states of similar models have been experimentally prepared in various quantum platforms\cite{Kokail2019S1810.03421,Zhang2021A2104.12636,Rosenberg2022E2106.01264,Zhang2022D,Xu2023C,Kim2023P2307.14627,Zheng2024T,Zhang2024Q2409.09729}, making them suitable for future experimental validation of our approach.

\textit{Model Architecture.} --- Our model employs a  transformer architecture\cite{NIPS2017_Transformer}, designed to model the conditional distribution $p_\text{mdl}(\vect{b}|\vect{P}, \vect{g})$ by leveraging its autoregressive, sequence-generation capabilities. Given the sequential nature of classical shadow data, a GPT model\cite{OpenAI2023G2303.08774} is well-suited for this task, as the model can condition on the prior measurement outcomes to progressively generate the next measurement outcome, akin to the next-token-prediction task in language modeling.

For each input data triple $(\vect{g}, \vect{P}, \vect{b})$, we embed each token as a vector in a $d$-dimensional representation space (with $d=128$) to standardize the format for processing by the transformer model. The real-valued Hamiltonian parameters $\vect{g}$ are all encoded into a single $d$-dimensional vector via a feed-forward neural network, as the orange triangle in \figref{fig:flow}. The observable sequence $\vect{P}$ and measurement outcome sequence $\vect{b}$ are interleaved and rearranged into a sequence of length $2N$, expressed as $(P_1, b_1, P_2, b_2, \dots, P_N, b_N)$, as shown in \figref{fig:flow}. This is to preserve the original spacial locality of measurement information within the sequence. Then each token in this rearranged sequence is embedded into a $d$-dimensional vector using a learnable token embedding layer, and each is supplemented with learnable positional encodings to encode absolute positional information, as the purple triangle in \figref{fig:flow}. Finally, the embedding vector of $\vect{g}$ is prepended to the $2N$ embedding vectors of $\vect{P}$ and $\vect{b}$, transforming the input data into $(2N + 1)$ $d$-dimensional embedding vectors, which serve as input to the transformer model.

The transformer architecture consists of a stack of decoder-only layers, each composed of a multi-head self-attention layer followed by a feed-forward neural network, with pre-norm layer normalization\cite{norm}. The sequence of embedding vectors is processed through these transformer layers, yielding a sequence of output vectors. The output vectors corresponding to each $P_i$ position are passed through a final feed-forward network, as the pink triangle in \figref{fig:flow}, to compute the conditional log-likelihood of the measurement outcomes $b_i$, specifically $\log p_\text{mdl}(b_i | P_{1:i}, b_{1:i-1}, \vect{g})$. This allows the total log likelihood of the measurement outcome sequence $\vect{b}$ to be expressed as
\begin{equation}
\log p_\text{mdl}(\vect{b} | \vect{P}, \vect{g}) = \sum_{i=1}^{N} \log p_\text{mdl}(b_i | P_{1:i}, b_{1:i-1}, \vect{g}).
\end{equation}
This enables us to train the model by calculating the loss function in \eqnref{eq: loss} and applying gradient backpropagation for optimization.

\begin{figure}[b]
\centering
\includegraphics[width=0.42\textwidth]{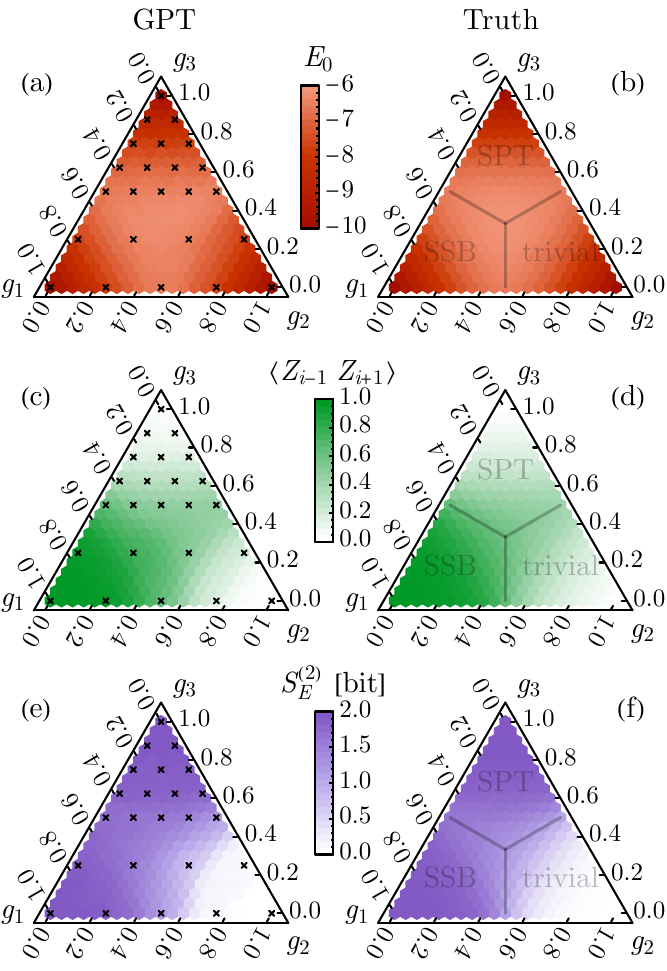}
\caption{GPT predictions of the cluster-Ising model ground state properties: (a) ground state energy $E_0$, (c) correlation function $\expval{Z_{i-1} Z_{i+1}}$ of order parameters, (e) 2nd Rényi entanglement entropy $S_E^{(2)}$ of a three-site reduced density matrix, all as a function of the Hamiltonian parameters $(g_1, g_2, g_3)$. Compare with the ground truth in (b), (d), (f) respectively. The model were only trained on a few parameter points marked out by black crosses.}
\label{fig:Z2Z2}
\end{figure}

\textit{Training and Prediction.} --- We train two separate ShadowGPT models on simulated data generated from above mentioned two families of Hamiltonians respectively, both with $N=10$ qubits. For the transverse-field Ising model, we sample totally $8\times 10^4$ classical shadows at $8$ parameter points $g=0,0.25,0.4,0.5,0.6,0.75,0.9,1$; while for the cluster-Ising model, we sample totally $24 \times 10^4$ classical shadows at $24$ parameter points $(g_1,g_2,g_3)$ within the simplex defined by $g_1+g_2+g_3=1$, whose positions are marked out by black crosses in \figref{fig:Z2Z2}(a). The models are trained using the AdamW optimizer\cite{AdamW_loshchilov2018decoupled} with a cosine annealing warm restart learning rate scheduler\cite{warmrestart_loshchilov2017sgdr}. 

The trained models can then be used to predict ground state physical properties using classical shadow reconstruction from the model-generated shadow data. Given a specified Hamiltonian parameters $\vect{g}$, we first sample a sequence of random Pauli observables $\vect{P}$ uniformly, and then generate the corresponding measurement outcomes $\vect{b}$ using the trained model $p_\text{mdl}(\vect{b}|\vect{P},\vect{g})$, as if the GPT model effectively simulates the quantum device behavior under randomized measurements. By repeating this process, we collect an ensemble of $(\vect{P}, \vect{b})$ data, from which we can estimate various physical properties of the ground state $\rho(\vect{g})$ using shadow tomography based on \eqnref{eq: shadow_tomo}. The median of means (MoM)\cite{MoM_doi:10.1137/1027074,MoM_JERRUM1986169} trick is used in estimating the ensemble average on the shadow data (with group number chosen to be $10$) to improve the estimation stability. In the following, we will evaluate the model performance by comparing the predicted properties with the ground truth values obtained from the exact solutions. 

\begin{figure}[b]
\centering
\includegraphics[width=0.42\textwidth]{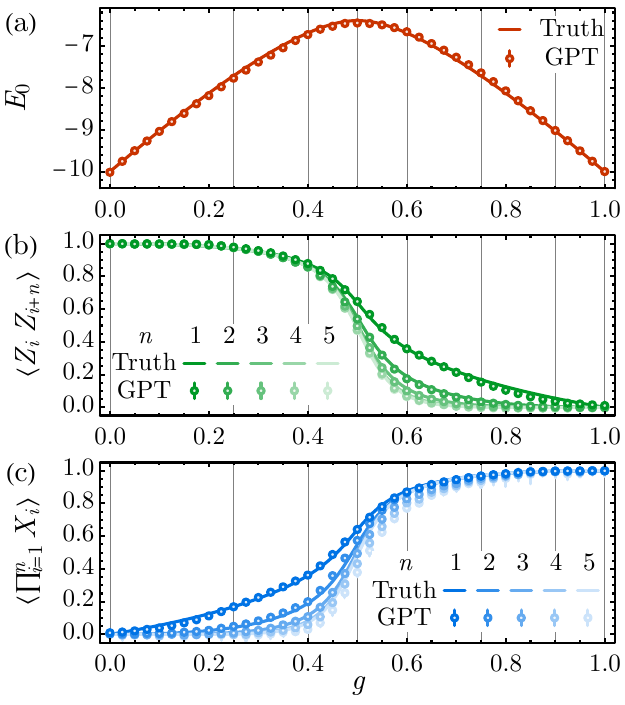}
\caption{GPT predictions of the transverse-field Ising model ground state properties: (a) ground state energy $E_0$, (b) correlation function $\expval{Z_i Z_{i+n}}$ of order parameters, (c) correlation function $\expval{\prod_{i=1}^{n}X_i}$ of disorder parameters (symmetry defects), all as a function of the Hamiltonian parameter $g$. The model were only trained on a few $g$ values indicated by the vertical gray lines.}
\label{fig:TFI}    
\end{figure}

\textit{Model Performance.} --- For the transverse field Ising model, the results in \figref{fig:TFI} include the model-predicted ground state energy $E_0$ and correlation functions of the \textit{order} and \textit{disorder} parameters, i.e., $\langle Z_i Z_{i+n} \rangle$ and $\langle \prod_{i=1}^n X_i \rangle$ respectively. At each parameter point $g$, the GPT model generates $3 \times 10^5$ classical shadows to make these predictions. Despite the training data being sampled only from a few sparse values of $g$, indicated by the vertical gray lines in \figref{fig:TFI}, the trained model interpolates predictions for previously unseen values of $g$ reasonably well.

The Kramers-Wannier duality ensures that the order and disorder parameter correlations must be related by $\langle Z_i Z_{i+n} \rangle (g) = \langle \prod_{i=1}^n X_i \rangle (1 - g)$. However, as seen in \figref{fig:TFI}(b) and (c), the prediction quality for the disorder parameter correlation (the $X$-string observable) is lower than for the order parameter correlation, particularly for large $n$. This discrepancy aligns with expectations for Pauli-based classical shadow tomography, as the strictly local and unentangled measurement scheme is less effective for estimating observables involving large operator sizes. Improvement could be achieved using finite-entanglement classical shadow tomography schemes\cite{Hu2022H2102.10132, Hu2023C2107.04817, Hu2022L2203.07263, Akhtar2023S2209.02093, Bertoni2024S2209.12924, Ippoliti2023O2212.11963, Ippoliti2024C2305.10723, Akhtar2024M2308.01653, Liu2023P2311.00695, Hu2024D2402.17911, Akhtar2024D2404.01068, Zhang2024H2406.11788}, though encoding such measurement observables presents a challenge that requires future research.

For the cluster-Ising model, \figref{fig:Z2Z2} presents the model-predicted ground state energy $E_0$, the order parameter correlation $\langle Z_{i-1} Z_{i+1} \rangle$, and the 2nd Rényi entanglement entropy $S_E^{(2)} = -\log(\Tr\rho_A^2)$ for a region $A$ containing three sites. At each parameter point $\vect{g}=(g_1, g_2, g_3)$, the GPT model generates $2 \times 10^5$ classical shadows for estimating the ground state energy and correlation function, and $3 \times 10^5$ for the Rényi entropy. Comparing the GPT predictions in \figref{fig:Z2Z2}(a), (c), and (e) with the ground truth in \figref{fig:Z2Z2}(b), (d), and (f), we observe strong predictive performance across the parameter space, even with training on limited parameter points. The ground state energies of the three cluster-Ising phases are identical, reflecting the triality. The SSB, trivial, and SPT phases are distinguished by the correlation function and Rényi entropy, though their phase boundaries are not sharp due to finite-size effects.

\textit{Conclusion.} --- We have introduced a novel shadow-based data-driven model, ShadowGPT, to address quantum many-body problems using classical machine learning. Our approach leverages the classical shadow tomography technique to generate classical shadow data from quantum experiments, which is then used to train a generative transformer model to predict ground state properties of quantum many-body systems. We have demonstrated the effectiveness of our model on two well-known quantum many-body models, the transverse-field Ising model and the cluster-Ising model, showing accurate predictions of ground state energy, correlation functions, and entanglement entropy. Our results opens up a promising avenue for utilizing quantum experimental data to solve quantum many-body problems, paving ways for building foundational models for future AI-for-Quantum applications.

The source code of this research work is available at the GitHub repository\cite{github}.

\begin{acknowledgments}
We thank Weijing Dai and Shuyi Li for helpful discussions. JY is supported by the Key-Area Research and Development Program of Guangdong Province Grant No.2020B0303010001 and No.2019ZT08X324. YZY is supported by a starup fund from the University of California at San Diego. We acknowledge OpenAI ChatGPT 4o for providing writing assistance. 
\end{acknowledgments}

\bibliography{ref}

\end{document}